# Giant nonlinear Hall effect driven by continuous Mott transition in twisted bilayer $WSe_2$

Meizhen Huang[1,3], Zefei Wu[1,2,3] ✉, Jinxin Hu[1,3], Xiangbin Cai[1], En Li[1], Liheng An[1], Xuemeng Feng[1], Ziqing Ye[1], Nian Lin[1], Kam Tuen Law[1] ✉, Ning Wang[1] ✉

[1]Department of Physics and Center for Quantum Materials, The Hong Kong University of Science and Technology, Hong Kong, China

[2]Present address: Department of Physics and Astronomy and National graphene Institute, University of Manchester, Manchester M13 9PL, UK.

[3]These authors contributed equally: Meizhen Huang, Zefei Wu, Jinxin Hu.

✉e-mail: phwang@ust.hk; phlaw@ust.hk; zefei.wu@manchester.ac.uk

## Abstract

**The recently discovered nonlinear Hall effect (NHE) in a few non-interacting systems provides a novel mechanism to generate second harmonic electrical Hall signals under time-reversal-symmetric conditions. Here, we introduce a new approach to engineering NHE by using twisted moiré structures. We find that the twisted $WSe_2$ bilayer exhibits a NHE when tuning the Fermi level to the moiré flat bands. Near half-filling of the first moiré band, the nonlinear Hall signal shows a sharp peak with the generation efficiency at least two orders of magnitude larger than those in previous experiments. We propose that the giant NHE and diverging generation efficiency originate from a mass-diverging type continuous Mott transition, which is evidenced by resistivity measurements. This work demonstrates not only how interaction effects can couple to Berry curvature dipoles to produce novel quantum phenomena, but also what NHE measurements can provide for developing a new tool to study the quantum criticality.**

## Main Text

The nonlinear Hall effect (NHE) can arise intrinsically from the Berry curvature dipole (BCD) moments of two-dimensional (2D) materials[1]. Unlike linear Hall effects, NHE possesses a transverse voltage oscillating at twice the frequency of the driving alternating current as well as a direct-current signal that is converted from the driving alternating-current. Materials with a tunable nonlinear Hall response can offer a broad range of technological applications that require second harmonic generation or rectification such as in efficient energy harvesting, next-generation wireless techniques, and infrared detectors[2, 3, 4]. The rectification and frequency doubling through the NHE are achieved by the intrinsic material property, and therefore do not have any thermal voltage thresholds and/or the transition time innate to semiconductor junctions/diodes[4, 5]. To date, however, experimental observations of the NHE are largely limited to a small class of non-centrosymmetric materials with non-zero Berry curvatures (BCs), such as $WTe_2$ (Ref. 6, 7), strained $MoS_2$ (Ref. 8) and corrugated graphene[9], in which the broken inversion and rotational symmetries allows the presence of finite BCDs[10, 11].

The development of van der Waals assembling techniques offer exciting opportunities to engineer heterostructures with exotic physical properties beyond those of individual materials[12, 13]. For example, when stacking two monolayer $WSe_2$ together with a small twist angle, the interlayer hybridization opens an energy bandgap at the band folding point and results in flat sub-bands[14, 15]. The modified band structures have led to the observation of emergent phenomena related to electron correlations beyond the expectations from single particle physics, such as continuous Mott transition and possible superconducting phases[16, 17, 18, 19]. In this work, we demonstrate that the twisting technique can provide an additional route to engineering the NHE. We report that electron correlation induced giant NHE happens near the half-filling of the twisted $WSe_2$ ($tWSe_2$) moiré bands. Near half-filling, we observe a giant second harmonic Hall voltage $V_{\perp}^{2\omega}$ of ~ 15 mV driven by a longitudinal voltage $V_{/\!/}^{\omega}$ of ~ 3.5 mV. The corresponding nonlinear Hall generation efficiency $\eta$ defined as $V_{\perp}^{2\omega}/\left(V_{/\!/}^{\omega}\right)^2$ achieves the level of $10^3$ $V^{-1}$, which is at least two orders of magnitude higher than the maximum value in non-twisted materials reported so far[9]. While the standard theory proposed by Sodemann and Fu can be used to understand the experimental data away from half-filling[1], the giant enhancement of the nonlinear Hall signal cannot be understood even when the moiré bands and strained induced BCD are taken into account. To understand the giant NHE, we notice that $\eta$ is proportional to the effective mass of the quasiparticles. We measured the temperature ($T$) dependence of the resistance and showed that there is a continuous Mott transition near half-filling. The transition gives rise to a divergent quasiparticle effective mass and enhances the NHE. This is the first demonstration of how the NHE can be dramatically enhanced near the quantum phase transition point. At the same time, the demonstration of the strongly enhanced $\eta$ provides a new experimental tool to understand the nature of the continuous metal-to-Mott-insulator transition which is one of the most outstanding and interesting problems in condensed matter physics[20, 21, 22, 23, 24].

## Nonlinear Hall signals in $tWSe_2$

Dual-gate devices composed of metal top gate/hBN/$tWSe_2$/bottom contacts/hBN are fabricated, as schematically shown in the inset of Fig. 1a. Here we use platinum as the bottom electrodes to ensure a good contact to the valence band of $WSe_2$. Fig. 1a shows the resistance $R$ of a typical high-quality $tWSe_2$ (twist angle $\theta = 2.0°$) as a function of filling $f$ (where $f = -1$ denotes there are two holes per moiré unit cell) at the temperature $T = 1.5$ K. The appearance of a resistance peak at the half-filling indicates the existence of strong correlation effects within the first moiré band, which is consistent with the results reported from other high-quality twisted moiré devices[12, 13, 16, 17].

It is important to note that unstrained $tWSe_2$ are not expected to exhibit NHE because of its three-fold rotational symmetry which forces the BCD to be zero. However, based on our scanning transmission electron microscopy (STEM) and scanning tunneling microscope (STM) characterization for $tWSe_2$ samples with $\theta$ ranging from 1° to 4°, we found that strain-induced three-fold rotational symmetry breaking universally exists in the fabricated devices (which is estimated to be in the range of 0.2~0.9%, see Supplementary Fig. 1)[25]. As illustrated in the calculation (shown in Fig. 1b-c), the strain reduces the symmetry of the $tWSe_2$ from $D_3$ to $C_1$, changes the BC distribution and induces non-zero BCD such that the NHE can occur (See Supplementary Fig. 2 for details)[26].

To measure the NHE, we apply a driving alternating-current $I^{\omega}$ with frequency $\omega$ between the source and drain (shown in the inset of Fig. 1d), and measure the Hall voltage $V_{\perp}^{2\omega}$ oscillating at $2\omega$. At half-filling of a $\theta = 2.0°$ sample, $V_{\perp}^{2\omega}$ increases nonlinearly and scales quadratically with $I^{\omega}$ (plotted by blue circles, Fig. 1d), exhibiting a clear nonlinear charge response. For comparison, we also collect the data of a monolayer $WSe_2$ device with the same device structure (enlarged by $2 \times 10^6$ times and plotted by black squares, Fig. 1d). In contrast, $V_{\perp}^{2\omega}$ for monolayer $WSe_2$ can hardly be observed at a similar carrier density $p \sim 1 \times 10^{12}$ cm$^{-2}$. In analogy to the spin Hall angle which measures the efficiency of the spin-charge conversion, the nonlinear Hall generation efficiency can be expressed as $\eta = V_{\perp}^{2\omega}/\left(V_{/\!/}^{\omega}\right)^2$. In $tWSe_2$, the observed nonlinear Hall generation efficiency $\eta \sim 10^3$ V$^{-1}$ is orders of magnitude higher than the values in monolayer $WSe_2$ sample ($\eta \sim 0.017$ V$^{-1}$) and other non-interacting systems (a comparison can be found in Supplementary Table 1)[6, 7, 9, 27, 28, 29].

## Giant NHE at the half-filling state

Since the nonlinear Hall signal maximizes when the applied current is parallel to the dipole[1], we use a device with disc geometry to study the angular and filling dependence of the NHE (details can be found in Supplementary Fig. 3). Fig. 2a shows the optical image of the disc-shaped sample. We apply a driving alternating-current $I^{\omega}$ between a pair of the twelve electrodes at angle $\delta$ from the zigzag direction of the $WSe_2$ crystal, and measure the second harmonic voltage $V_{\perp}^{2\omega}$. Polar contour plot of $V_{\perp}^{2\omega}$ is shown in Fig. 2b, where the angular axis denotes the current injection direction and the radial axis denotes the filling. The angular and filling dependent $V_{\perp}^{2\omega}$ exhibits two features. Firstly, $V_{\perp}^{2\omega}$ switches signs when the current direction and the voltage probe connection are reversed simultaneously, indicate a BCD induced nonlinear charge response (also can be seen in Fig. 2c). Secondly, $V_{\perp}^{2\omega}$ near the half-filling of the moiré bands is orders of magnitude higher than other fillings (marked by the black circles in Fig. 2b and also can be seen in Fig. 2e).

To be more specific, $V_{\perp}^{2\omega}$ versus $I^{\omega}$ at different filling and $V_{\perp}^{2\omega}$ versus $f$ along $\delta = 81.5°$ is plotted in Fig. 2d and Fig. 2e respectively. As can be seen in Fig. 2d, the $V_{\perp}^{2\omega} - I^{\omega}$ characteristic is robust at all filling. $V_{\perp}^{2\omega}$ increases nonlinearly and scales quadratically with $I^{\omega}$, exhibiting a clear NHE. While the quadratic $V_{\perp}^{2\omega} - I^{\omega}$ characteristic is robust at all filling, the $V_{\perp}^{2\omega}$ signal depends strongly on the filling. As shown in Fig. 2e, $V_{\perp}^{2\omega}$ shows a sharp peak at half-filling and drops by three orders of magnitude slightly away from the half-filling. Though the sharp peak at half-filling can be observed along most of the current injection directions and the sample inhomogeneity may play a role in a few directions, its unusual sensitivity to filling suggests that the strong correlation effects at the half-filling may play an important role in generating the giant nonlinear response.

To understand the origin of the giant NHE, we note that besides the strong peak near half-filling, as shown in the upper panel of Fig. 2f, the nonlinear Hall signal shows a small peak when the filling $f$ is near −0.63. In the non-interacting case and small twist angle of $\theta$ = 2.0°, the band structure under strain as well as the BCD of the top moiré bands can be rather accurately described by the continuum model (See Supplementary Note 1 for details). From the lower panel of Fig. 2f, we can clear see that the calculated Berry curvature dipole $D$ shows a peak at $f$ = −0.63, which is consistent with the experimental measurement (Details can be found in Supplementary Fig. 4). The model can also capture the sign change of the nonlinear Hall signal near $f$ = −0.30/−0.77. When the filling $f$ is near 0, the Berry curvature dipole is almost zero as observed in the experiment. However, the non-interacting electron model cannot explain the giant NHE near $f$ = −0.5. Since the large enhancement of the nonlinear Hall signals happens near half-filling, it is natural to deduce that interacting effects are playing an important role.

At $\theta$ = 2.0°, we can estimate that the on-site Coulomb repulsion energy $U$ of each site is about 10.5 meV (details of calculation can be found in Methods). Our calculations show the band width $W_b$ of the first hole moiré band for $\theta$ = 2.0° is ~10 meV (Ref. 26). The large ratio of $U/W_b$ suggests the possibility of a metal-insulator transition near the half-filling of the band[30]. Mott proposed a carrier-density-vanishing type first-order metal-insulator transition mechanism[31]. In contrast, recent experiments in heterobilayer $MoTe_2/WSe_2$ and $tWSe_2$ support the presence of a mass-diverging type second-order metal-insulator transition[18, 19]. Importantly, the nonlinear Hall signal is proportional to both $D$ and the effective mass $m^*$ such that $V_{\perp}^{2\omega} \propto Dm^*$. Therefore, a diverging effective mass can strongly enhance the nonlinear Hall signal as observed experimentally. In the following, we demonstrate the continuous metal-to-Mott-insulator transition by resistivity measurements.

## Mass-diverging type continuous Mott transition

To verify the continuous metal-insulator transition, we study the phase diagram of the correlated insulating and metallic regions in our $tWSe_2$ sample which is identified through the temperature dependence of the resistance as shown in Fig. 3a and Fig. 3b. In the region of $-0.28 < f < -0.25$, where $f$ denotes the filling, the temperature dependent resistivity can be express up to 15 K by the relation $\rho = \rho_0 + \alpha T^2$, where $\rho_0$ is the residue resistivity and $\alpha^{1/2} \propto m^*$ is the fitting parameter, showing a Fermi liquid behavior. The parameter $\alpha^{1/2}$ is enhanced as $f$ approaches the quantum criticality from the metallic side (See Supplementary Fig. 5). In contrast, for a range of fillings near the correlated insulator, $T$-linear behavior is observed to the lowest temperatures. Fig. 3b plotted the $R$-$T$ curves near the metal-insulator transition point. The resistivity shows a $T$-linear behavior ($\rho = \rho_0 + \beta T$) at $-0.35 < f < -0.28$ at temperature $T < 10$ K and saturates at high temperatures. Upon increasing the doping, a correlated insulating phase with $\mathrm{d}\rho/\mathrm{d}T < 0$ is observed. Further increasing the doping, $T$-linear behavior is recovered to lowest temperatures. This 'anomalous metallic' behavior is similar to those in some strongly correlated cuprates[20], magic-angle twisted bilayer graphene[32], and recent works in twisted transition metal dichalcogenides[18, 19]. As shown in Fig. 3c, the extracted insulating gap as a function of filling drops smoothly from the highest value at the half-filling to nearly zero at the quantum critical point, indicating a continuous phase transition. Similar behavior is also observed when tuning the displacement field (Fig. 3d), in which the size of the gap can be tuned by 30%, mainly due to the limited tuning ability of the gate voltages (See supplementary Fig. 6 for the displacement electric field's effect).

## Giant NHE driven by continuous Mott transition

It is important to note that the moiré lattice of the twisted bilayer $WSe_2$ sample is triangular and such a system is supposed to be described by a Hubbard model with triangular lattices. It has been predicted that such a Hubbard model supports a continuous metal-to-Mott-insulator transition[33, 34, 35]. Near the transition, the quasiparticle effective mass diverges[21, 22]. While the quasiparticle properties are strongly modified, we expect that the transport properties of the quasiparticles can still be described by the Boltzmann equation. As a result, the nonlinear Hall signal is proportional to the effective mass as well as the Berry curvature dipole $V_{\perp}^{2\omega} \propto Dm^{*}$ and the nonlinear Hall generation efficiency $\eta = V_{\perp}^{2\omega}/\left(V_{/\!/}^{\omega}\right)^{2}$ also diverges as the effective mass diverges (See Supplementary Fig. 4 for details). As shown in Fig. 4a and Fig. 4b, $\eta$ extracted experimentally near the half-filling can be well fitted by using $m^{*}/m_{e} \propto \ln\frac{1}{|1-2f|}$, where $m_{e}$ is the bare electron mass[21]. Therefore, we demonstrated that the measurement of the diverging $\eta$ provides a new and very valuable tool to study the diverging behaviour of the effective mass near the critical regime of the metal-to-Mott-insulator transition.

At last, to highlight the twisted system as an ideal platform for generating nonlinear Hall signals, we compare the nonlinear Hall generation efficiency in various kinds of 2D materials in Fig. 4c (Supplementary Fig. 7-8 and Ref. 6, 7, 9, 27, 28, 29). Among them, $tWSe_2$ has the highest nonlinear Hall generation efficiency, which is 2-3 orders of magnitudes higher than those in the non-interacting systems. The fact that the twisted systems can possess a large nonlinear Hall signal with tunable amplitude is an important addition to the arsenal of van der Waals technologies. It can be used, for example, in efficient energy harvesting, next-generation wireless techniques, infrared detectors, and other applications which need second harmonic generation or rectification. Our work is also important for paving an additional route to generating and engineering the BCDs, which can be applied to other 2D materials with non-zero Berry curvatures.

## Conclusions

We highlight $tWSe_2$ as a highly tunable and correlated system for introducing and manipulating nonlinear Hall effects. We experimentally demonstrated the strong enhancement of the nonlinear Hall signal near the critical regime of the continuous metal-to-Mott-insulator transition. The measurement of the diverging behavior of the nonlinear Hall generation efficiency provides a new tool to study the novel properties of the critical regime.

## Methods

### Device fabrication

The transport devices in this study was fabricated using the tear and stack method. The hBN (~30 nm) and $WSe_2$ flakes are exfoliated on $SiO_2$/Si substrates and identified optically. Pt electrodes (10 nm) are patterned on bottom hBN. The top hBN (~15-30 nm) is picked up from the $SiO_2$/Si substrate using a polycarbonate/polydimethylsiloxane (PC/PDMS) stamp on a glass slide. We manually rotate the transfer stage by a small angle (1 ~ 4°) between picking up two $WSe_2$ pieces[36]. The entire top hBN/t$WSe_2$ stack is transferred onto the pre-patterned Pt electrodes placing on the bottom hBN. After that, a Ti/Au (10/30 nm) top gate is patterned.

### Transport measurement

Transport measurements were performed in cryogenic system which provides stable temperatures ranging from 1.4 to 300 K and fields up to 14 Tesla. The DC top gate was applied through our home-made DAC with resolution of 1 mV. AC bias voltage was applied to the source probe through SRS DS360. First and second harmonic signals were measured by Signal Recovery 7280 lock-in amplifier (impedance 100 MΩ).

### Carrier density, displacement field and twist angle determination

The carrier density $p$ can be determined by a combination of the Hall measurement and carrier density induced by $V_{\mathrm{tg}}$ and $V_{\mathrm{bg}}$ based on a capacitor model: $pe = C_{tg}\left(V_{\mathrm{tg}} - V_0\right) + C_{bg}V_{\mathrm{bg}}$, where $C_{tg}$ and $C_{bg}$ is the capacitance of the top BN and the bottom $SiO_2$, $V_0$ is the threshold voltage at which the Fermi level touches the flat band edge. The displacement field can be calculated by $D_{\mathrm{field}} = \left(C_{bg}V_{\mathrm{bg}} - C_{tg}V_{\mathrm{tg}}\right)/2\varepsilon_0$, where $\varepsilon_0$ denotes the vacuum dielectric constant. The twist angle is estimated from $p_0 = 2/\left(\frac{\sqrt{3}}{2}\lambda^2\right)$ and $\lambda = a/\left(2\sin\frac{\theta}{2}\right)$ where $p_0$ is the carrier density at the full-filling, and $a = 0.329$ nm is the lattice constant of $WSe_2$.

### Calculation of the on-site Coulomb energy

The on-site Coulomb energy $U$ of each site is estimated to be $e^2/(4\pi\varepsilon d_0)$, where $e$ is the elementary charge, $\varepsilon$ is the effective in-plane dielectric constant including screening and $d_0$ is the effective distance between each site[37]. Considering $d_0$ is 40% of the moiré wavelength and $\varepsilon = 3\varepsilon_0$, we get $U = e^2\theta/(4\pi\varepsilon_0 a\kappa)$, where $\kappa = 0.4\times3 = 1.2$ and $a = 0.33$ nm is the $WSe_2$ lattice constant.

## Acknowledgements

We thank Prof. Inti Sodemann and Oles Matsyshyn for insightful advices. We also thank Prof. Z. Y. Yang and Dr. Benjamin T. Zhou for fruitful discussions on experiments and theoretical analysis. Great supports from the National Key R&D Program of China (2020YFA 0309600) and the Research Grants Council (RGC) of Hong Kong (Project No. 16300717, C6025-19G and C7036-17W) are acknowledged. E.L. and N.L. thank the RGC funding of Hong Kong (CRF No. C6012-17E). Device fabrication was performed at the MCPF and WMINST of HKUST with great technical support from Ms. Wing Ki Wong and Dr. Yuan Cai.

## Author contributions

Z.-F.W. and M.-Z.H. conceived the project and designed the experiments. M.-Z.H. and Z.-F.W. fabricated the devices, performed the measurements and analyzed the data. X.-B.C. performed STEM characterizations. J.-X.H. performed theoretical computations under the supervision of K.-T.L. E.L. performed STM characterizations under the supervision of N.L. L.-H.A., X.-M.F. and Z.-Q.Y. provided technical support in the device fabrication process. Z.-F.W. and M.-Z.H. wrote the manuscript. N.W. and K.-T.L. supervised the work and polished the manuscript.

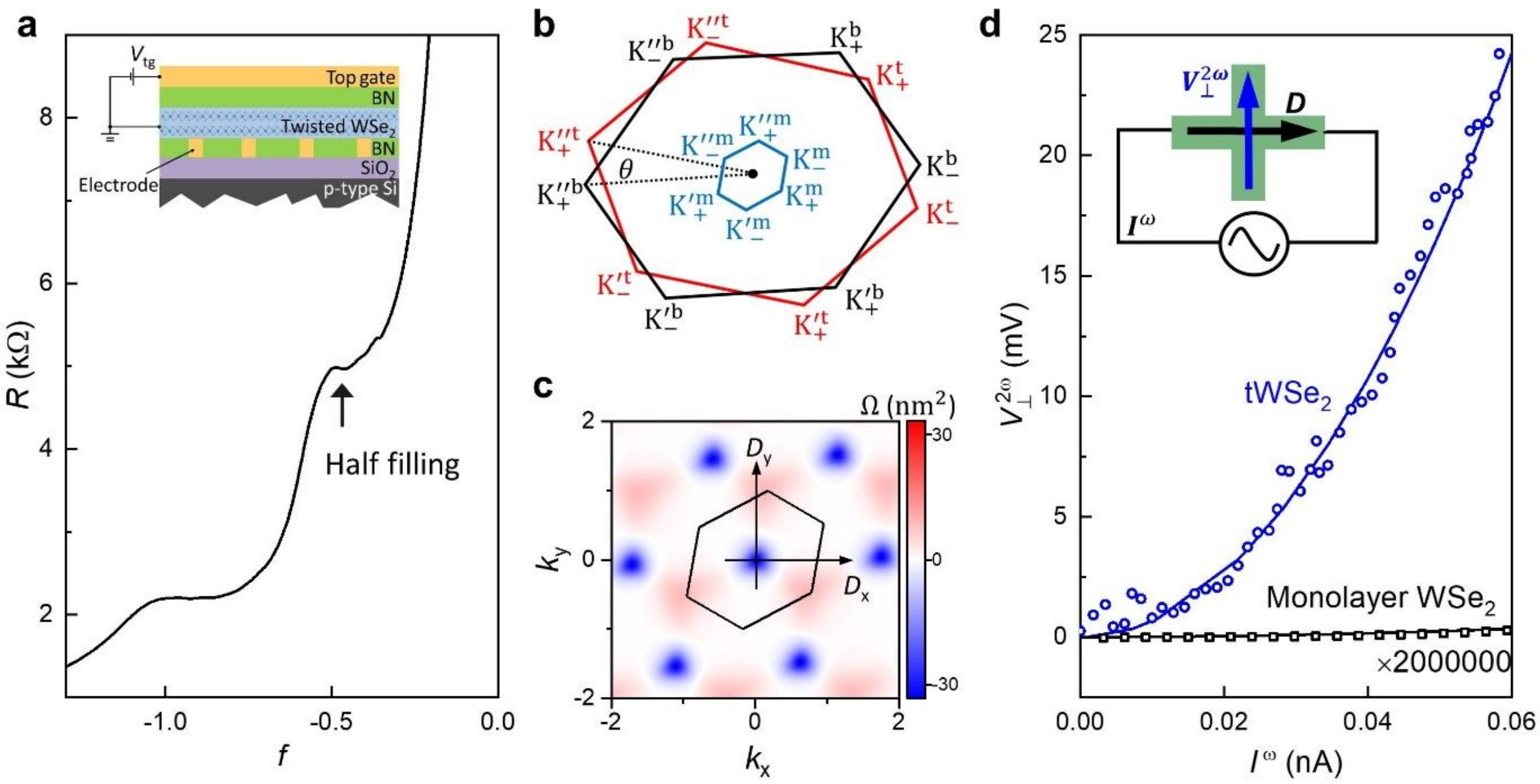

**Fig.1 | Basic characterization and strain induced berry curvature dipole of $tWSe_2$. a,** *R* plotted as a function of *f* for a standard $tWSe_2$ sample measured at $T$ =1.5 K. Inset: Schematic of the device structure. **b,** First Brillouin zone of the top layer (black), bottom layer (red) $WSe_2$ and the moiré Brillouin zone (blue) with strain. **c,** BC of the top moiré valence band with a strain strength of 0.6% along zigzag direction. The unbalanced BC distribution results in finite dipole along both zigzag (*x*) and armchair (*y*) direction. **d,** Blue: $V_{\perp}^{2\omega}$ as a function of $I^{\omega}$ measured at half-filling of $tWSe_2$ at $T$ =1.5 K. Black: $V_{\perp}^{2\omega}$ measured in monolayer $WSe_2$. The dots are experimental data and the solid lines are parabolic fits of the data. Inset: Illustration of the NHE. A driving alternating-current parallel to the dipole (*D*) can generate a second harmonic nonlinear Hall voltage.

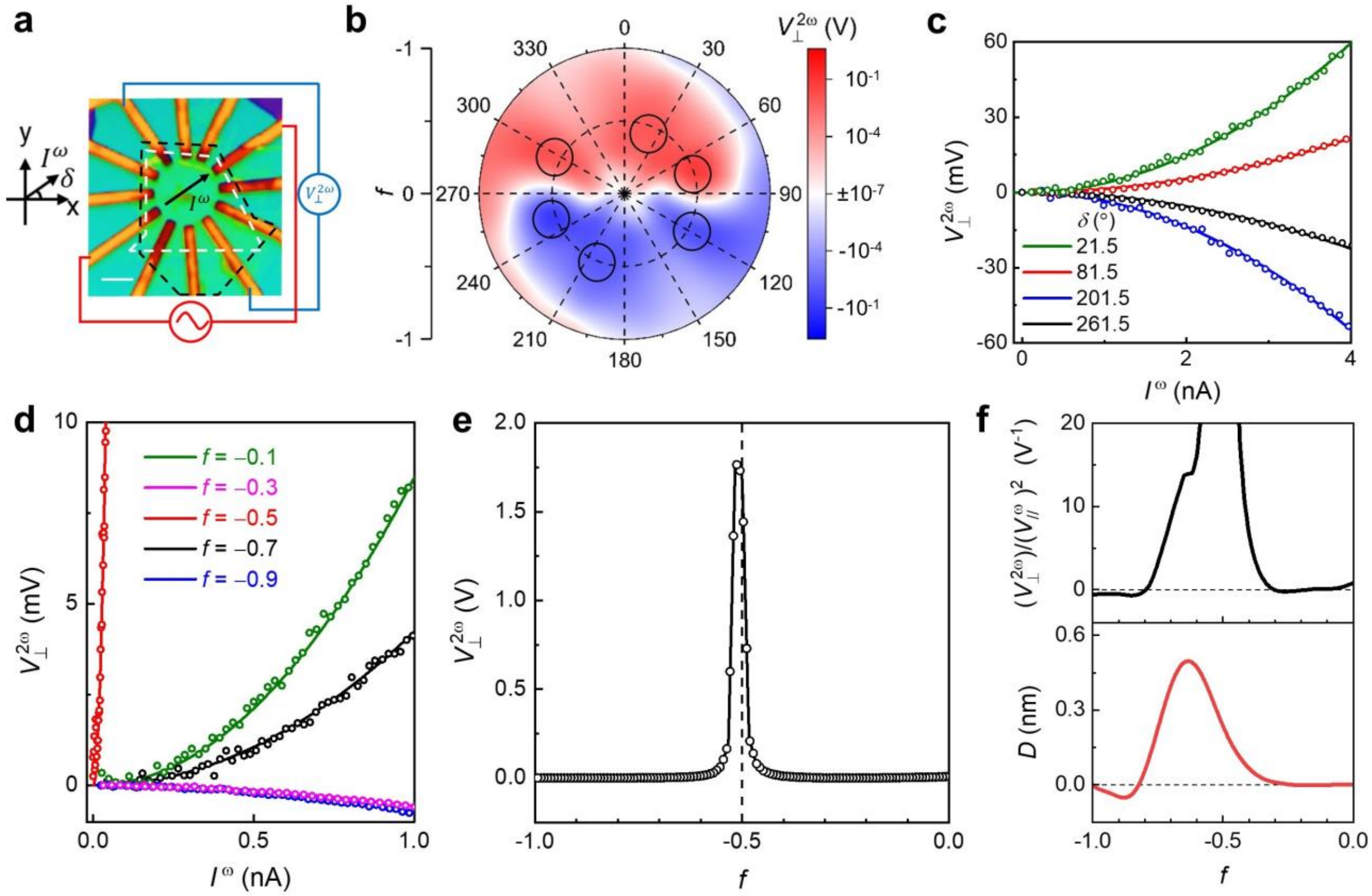

**Fig.2 | Giant nonlinear Hall effect in tWSe$_2$. a,** Optical image of disc-shaped sample (scale bar 4 μm). Black (white) dashed area corresponds to bottom (top) WSe$_2$ layer. *x* is the zigzag direction and *y* is the armchair direction of the WSe$_2$ crystal. $I^{\omega}$ is injected at angle $\delta$ from the zigzag direction. **b,** Angular and filling dependent $V_{\perp}^{2\omega}$ measured at $I^{\omega} = 0.5$ nA at $T$ = 1.5 K. **c,** $V_{\perp}^{2\omega}$ versus $I^{\omega}$ measured along different directions at $f$ = –0.74 at $T$ = 1.5 K. **d,** $V_{\perp}^{2\omega}$ versus $I^{\omega}$ measured at different fillings along $\delta = 81.5°$. The red curve is the same one as the blue curve shown in Fig. 1d. The dots are experimental data and the solid lines are parabolic fits of the data. **e,** $V_{\perp}^{2\omega}$ along $\delta = 81.5°$. A sharp peak is observed near the half-filling. **f,** $V_{\perp}^{2\omega}/\left(V_{/\!/}^{\omega}\right)^2$ extracted from experiment (upper panel) compared with *D* calculated from theory (bottom panel) along $\delta = 81.5°$.

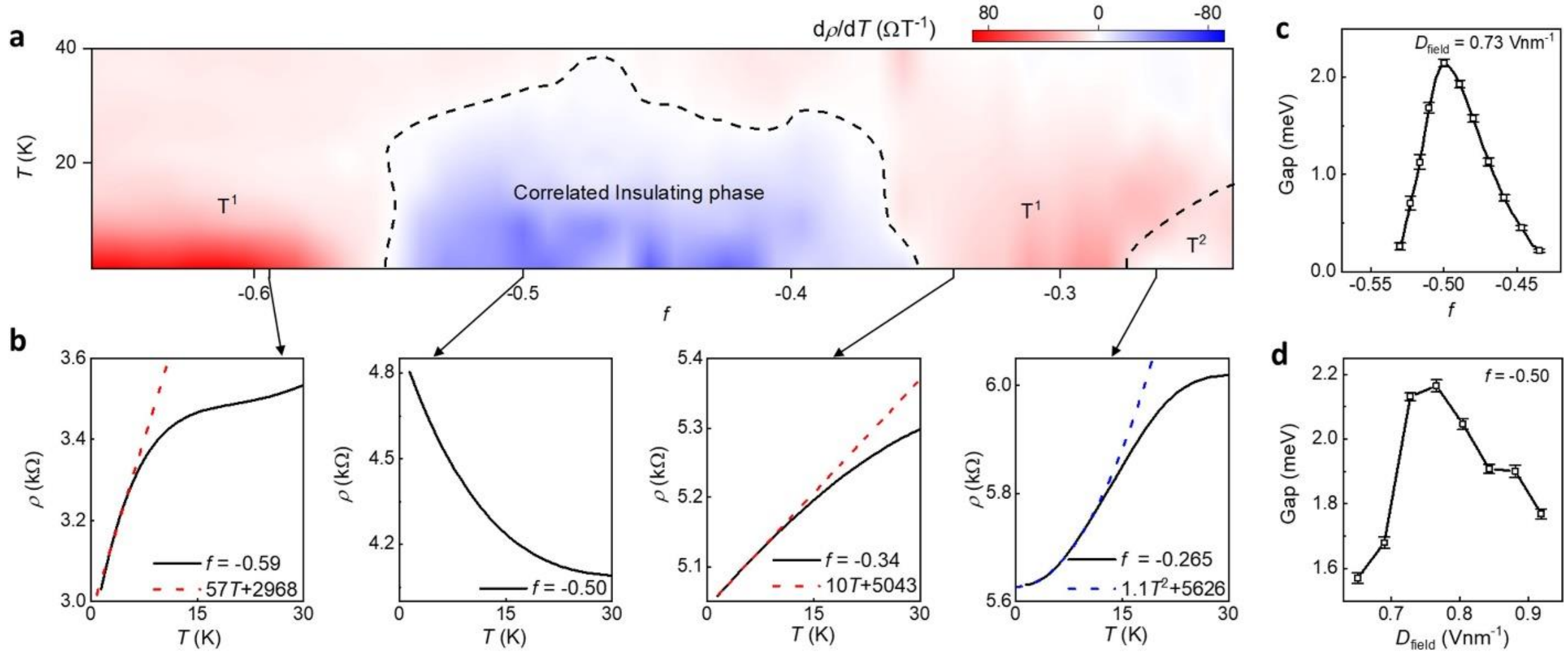

**Fig. 3 | Mass-diverging type continuous metal-insulator transition. a,** First derivative in temperature of the resistivity. **b,** Plots of resistivity (black) versus $T$ as well as linear (red) and quadratic (blue) fits at different filling. **c,** Plot of the measured insulating gap versus filling. **d,** Plot of the measured insulating gap versus displacement field.

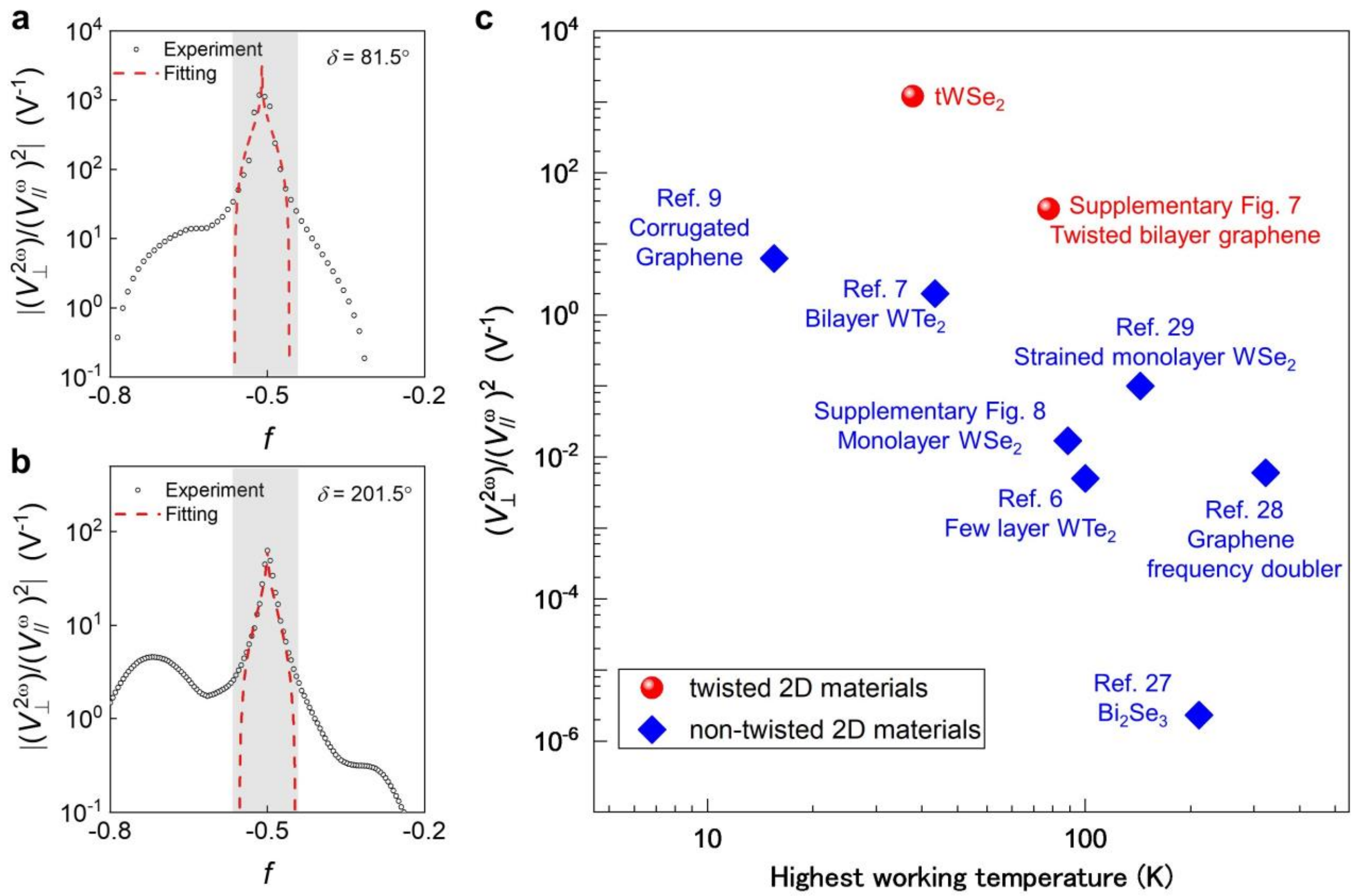

**Fig. 4 | Giant nonlinear Hall response from divergent effective mass. a, b,** Black dots: $V_{\perp}^{2\omega}/\left(V_{/\!/}^{\omega}\right)^2$ obtained along $\delta = 81.5°$ (**a**) and $\delta = 201.5°$ (**b**) from experiment near the half-filling. Red dashed line: theoretical fitting using the effective mass divergence formula. The grey area corresponds to the correlated insulating phase. **c,** Logarithmic plot of $V_{\perp}^{2\omega}/\left(V_{/\!/}^{\omega}\right)^2$ versus the highest working temperature for time-reversal-symmetric frequency doublers in various kinds of 2D materials.

Supplementary Information

for

# Giant nonlinear Hall effect driven by continuous Mott transition in twisted bilayer $WSe_2$

Meizhen Huang[1,3], Zefei Wu[1,2,3] ✉, Jinxin Hu[1,3], Xiangbin Cai[1], En Li[1], Liheng An[1], Xuemeng Feng[1], Ziqing Ye[1], Nian Lin[1], Kam Tuen Law[1] ✉, Ning Wang[1] ✉

[1]Department of Physics and Center for Quantum Materials, The Hong Kong University of Science and Technology, Hong Kong, China

[2]Present address: National graphene Institute, department of Physics and Astronomy, University of Manchester, Manchester M13 9PL, UK.

[3]These authors contributed equally: Meizhen Huang, Zefei Wu, Jinxin Hu.

✉e-mail: phwang@ust.hk; phlaw@ust.hk; zefei.wu@manchester.ac.uk

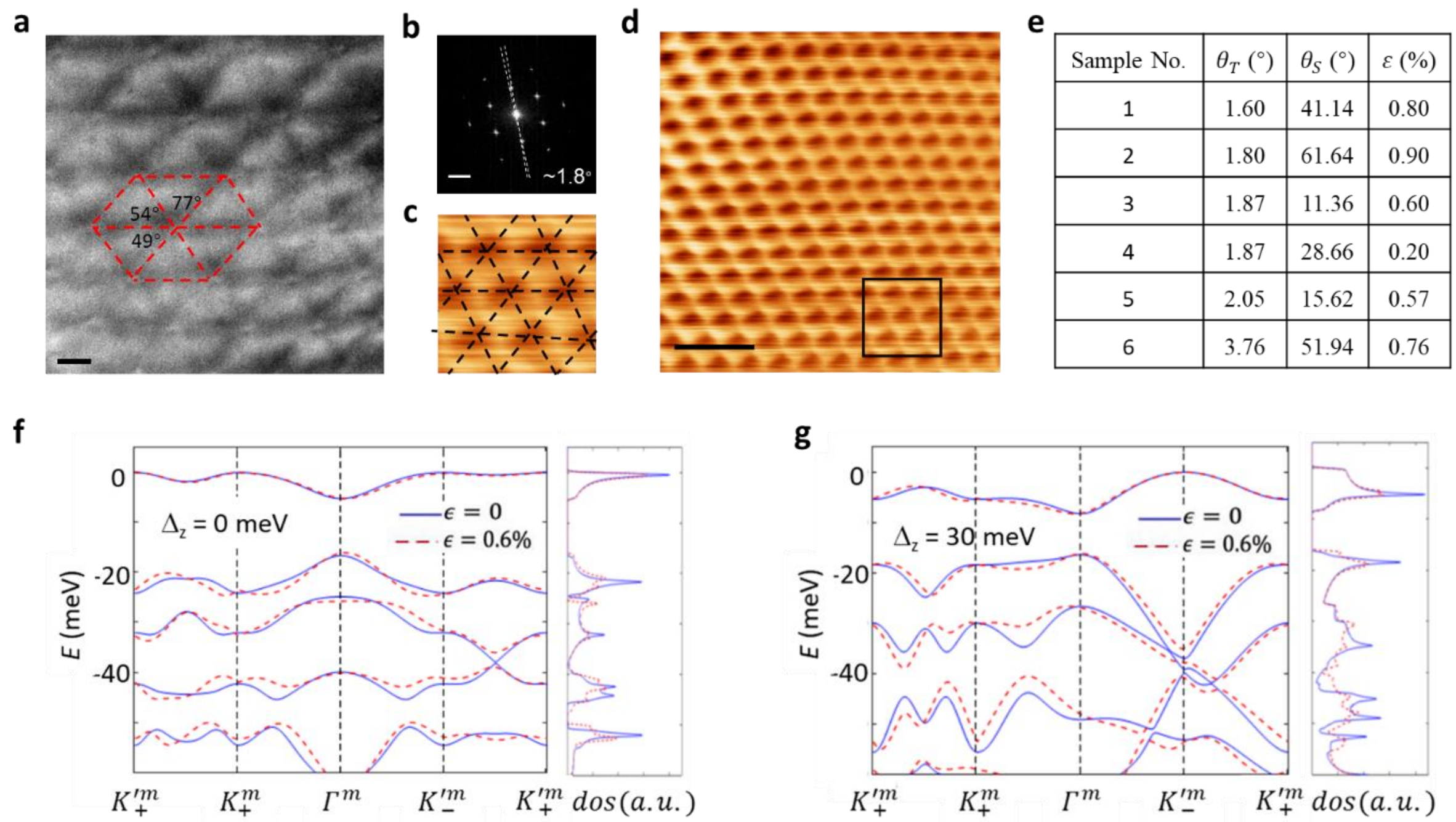

| Sample No. | $\theta_T$ (°) | $\theta_S$ (°) | $\varepsilon$ (%) |
|---|---|---|---|
| 1 | 1.60 | 41.14 | 0.80 |
| 2 | 1.80 | 61.64 | 0.90 |
| 3 | 1.87 | 11.36 | 0.60 |
| 4 | 1.87 | 28.66 | 0.20 |
| 5 | 2.05 | 15.62 | 0.57 |
| 6 | 3.76 | 51.94 | 0.76 |

**Supplementary Fig. 1 | Symmetry breaking effects in $tWSe_2$ sample. a,** High-resolution STEM image of the moiré superlattice formed from $tWSe_2$ (scale bar 5 nm). AA stacking points (W atom on W atom or Se atom on Se atom) are connected through red lines. The moiré superlattice exhibits strong distortion as outlined by the red dashed lines connecting the nearest AA stacking points with angles of 77°, 54° and 49°, largely deviating from the ideal angle of 60°. Using the uniaxial heterostrain model[1], the strain strength is calculated to be 0.90%. **b,** Twist angle (~1.8°) for the STEM sample can be determined by measuring the moiré lattice scale and the angular difference between the two sets of electron diffraction patterns. (scale bar 2.3 $nm^{-1}$). **c,** Enlarged image of the region marked by the dark square in (**d**) (area length 19.2 nm). **d,** STM image from $tWSe_2$ sample (scale bar 20 nm). The moiré superlattice is highly distorted and shows inhomogeneous strain strength. The strain strength is calculated to be 0.57% in this sample with a twist angle of 2.05°. e, Strain calculated in different STEM and STM samples. We conclude that significant symmetry breaking effects universally exist in $tWSe_2$. **f,g,** Band structure and DOS calculated at $\Delta_z$ = 0 meV (**f**) and $\Delta_z$ = 30 meV (**g**), the later staggered potential corresponds to half-filling ($f$ = –0.5) in Fig. 2e.

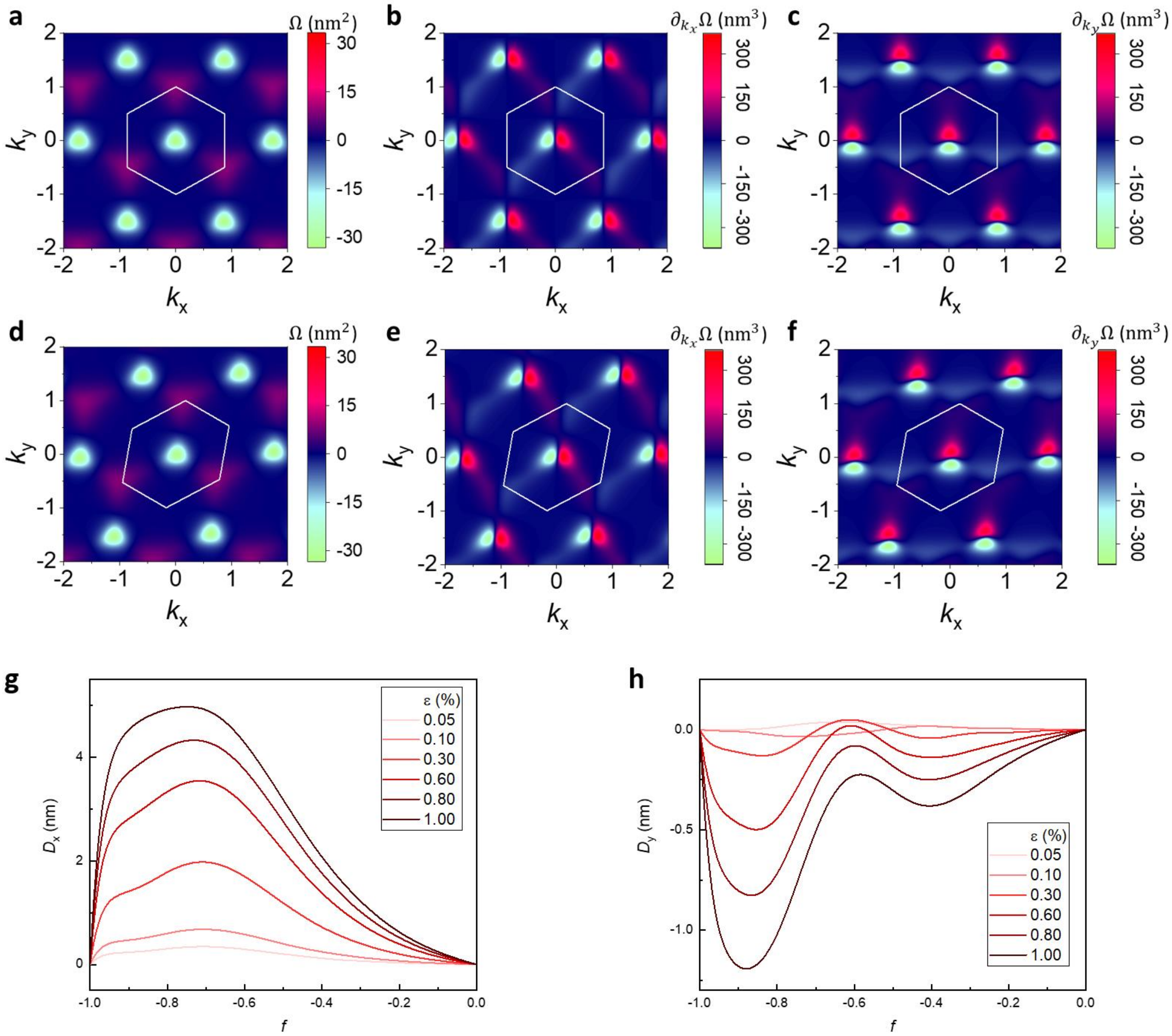

**Supplementary Fig. 2 | BC and BCD in tWSe$_2$. a,** The Berry curvature of the top moiré valence band without strain. **b**-**c**, Dipole density along zigzag $k_\mathrm{x}$ (**b**) and armchair $k_\mathrm{y}$ (**c**) direction without strain. The BCD, which measures the gain in total BC flux, vanish in this twist systems with $C_3$ symmetry. $k_\mathrm{x}$ and $k_\mathrm{y}$ are in units of nm$^{-1}$. **d-f,** The Berry curvature (**d**), dipole density along $k_\mathrm{x}$ (**e**) and $k_\mathrm{y}$ (**f**) direction after introducing a strain strength of 0.6% along zigzag direction. The unbalanced BCD distribution is displayed in (**e**)/(**f**) along $k_\mathrm{x}/k_\mathrm{y}$ direction, respectively. This breaking of three-fold rotational symmetry results in finite BCD. The staggered layer potential is 30 meV in the calculation, which corresponding to half-filling ($f$ = –0.5) in Fig. 2e. **g,h,** Theoretical values of $D_x$ (**g**) and $D_y$ (**h**) plotted as a function of filling based on a non-interacting model. $D_x$($D_y$) is the BCD along the zigzag(armchair) direction. The giant enhancement of the nonlinear Hall signal at half-filling is irrelevant to the strain strength since $D$ shows same shape with a smooth amplitude change for different strain strength.

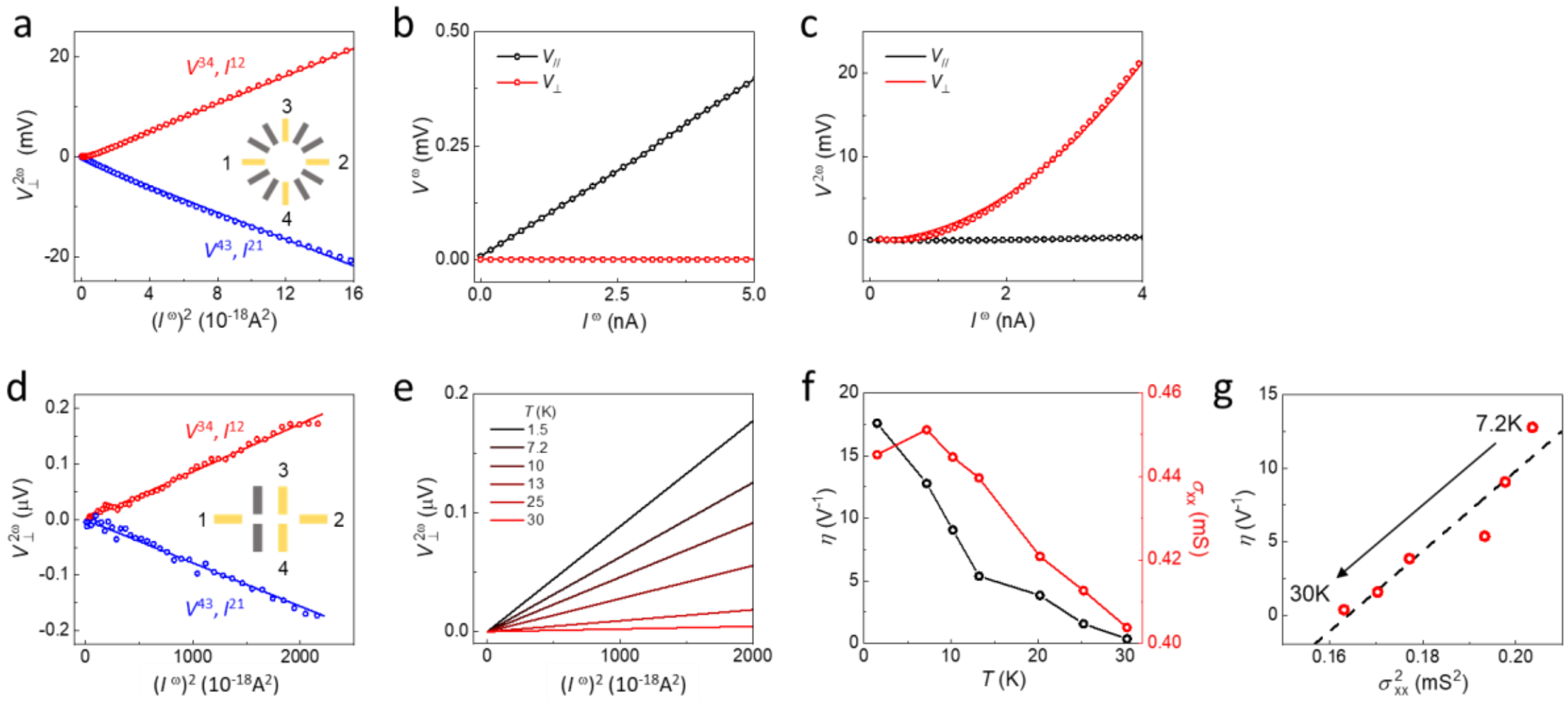

**Supplementary Fig. 3 | Additional NHE data of tWSe₂. a,** $V_{\perp}^{2w}$ measured at $T$ = 1.5 K at $f$ = –0.74 from different combinations of electrodes as functions of $(I^{\omega})^2$. Dots are measured value and lines are fitted curves. $V^{43}$ and $I^{12}$ indicate that the second harmonic voltage is measured through probes 4 and 3 when current $I^{\omega}$ is injected through probes 1 and 2. $V_{\perp}^{2\omega}$ scales quadratically with $I^{\omega}$ and changes its signs when both the voltage detection and current injection directions are reversed. Inset: electrode configuration of the device. **b,** The first harmonic longitudinal voltage $V_{/\!/}^{\omega}$ measured at $T$ = 1.5 K at $f$ = –0.74 increases linearly with current $I^{\omega}$. By correcting the misalignment of the electrodes, nearly zero $V_{\perp}^{\omega}$ is obtained, suggesting that the device is a non-magnetic system with protected time-reversal symmetry. **c,** A second harmonic Hall response $V_{\perp}^{2\omega}$ measured at $T$ = 1.5 K at $f$ = –0.74 dominates over the longitudinal response $V_{/\!/}^{2\omega}$. Dots are measured value and lines are parabolic fits of the data. All the features above verify the BCD-induced NHE in tWSe₂ and excluding other possible effects such as the asymmetric flake shape, side-jump and skew-scattering, contact junction, and thermoelectric effects[2].

(**d**)-(**g**) are the NHE data from another tWSe₂ sample ($\theta$ = 2.0°). **d,** $V_{\perp}^{2w}$ measured at $T$ = 1.5 K from different combinations of electrodes as functions of $(I^{\omega})^2$. Inset: electrode configuration of the device. **e**, $V_{\perp}^{2\omega}$ measured at different temperatures as a function of $(I^{\omega})^2$. **f,** $\eta$ (left, black) and $\sigma_{xx}$ (right, red) as a function of $T$. **g,** $\eta$ as a function of $(\sigma_{xx})^2$ at different temperatures. The dashed line is a linear fitting of experimental data. Though the linear fitting cannot rule out the possibility from other second harmonic transport mechanisms[3], the conclusion that the BCD dominates in our samples is solid based on the Hall dominated second harmonic response.

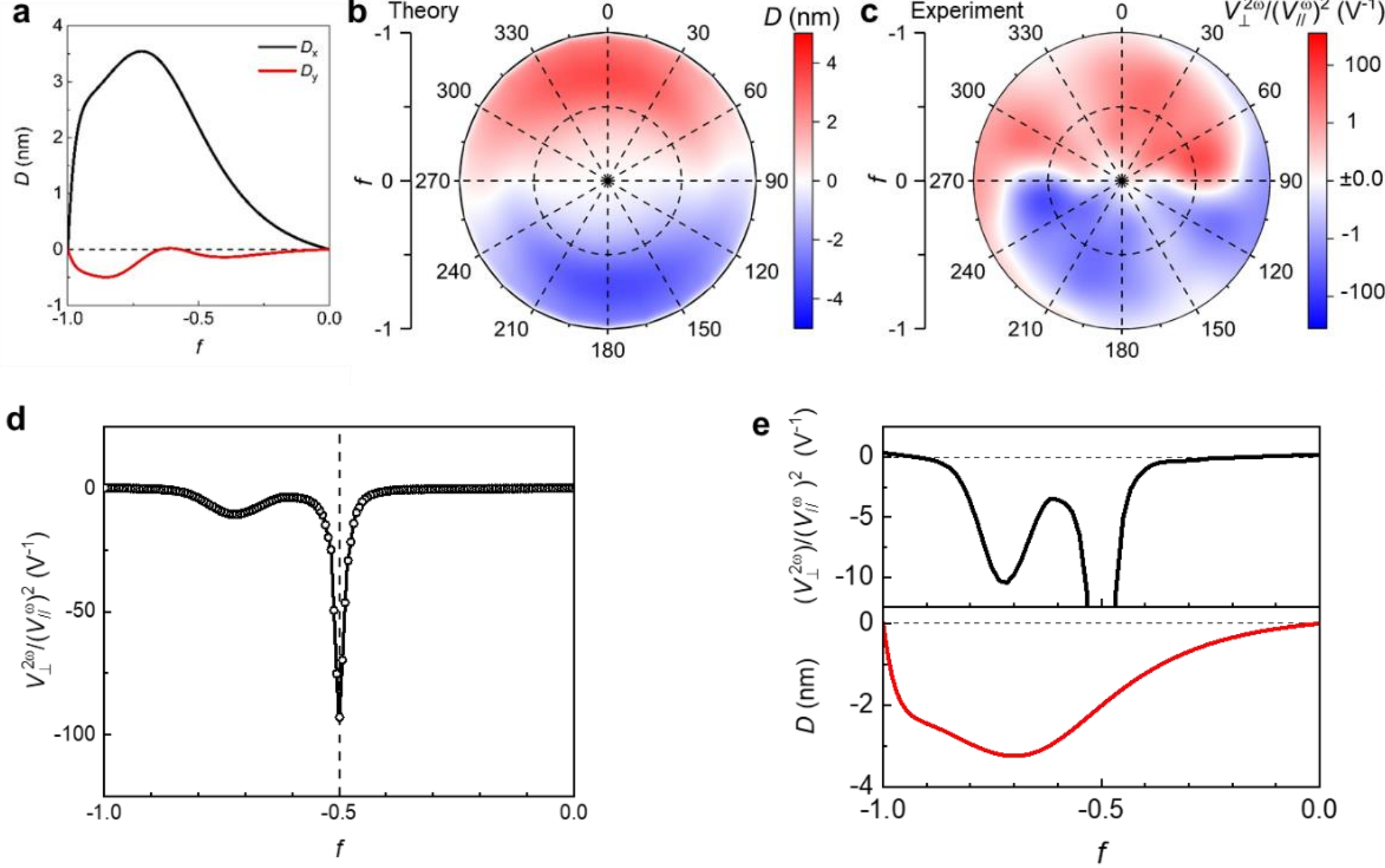

**Supplementary Fig. 4 | Theoretical calculation of the BCD and the NHE data at $\delta = 201.5°$. a,** Calculated BCD along both zigzag ($D_x$) and armchair ($D_y$) directions using a continuum model. **b,** Angular and filling dependences of theoretical calculated $D = D_x cos\delta + D_y \sin\delta$ if we assume that the sample resistance is isotropic[4]. **c,** Angular and filling dependent $V_\perp^{2\omega}/\left(V_{/\!/}^{\omega}\right)^2$ measured at $T = 1.5$ K. **d,** $V_\perp^{2\omega}/\left(V_{/\!/}^{\omega}\right)^2$ extracted from experimental data along $\delta = 201.5°$. A sharp peak is observed near the half-filling. **e,** $V_\perp^{2\omega}/\left(V_{/\!/}^{\omega}\right)^2$ extracted from experiment (upper panel) compared with $D$ calculated from theory (bottom panel) along $\delta = 201.5°$. A small peak appears near filling ~ –0.73, which is consistent with the theoretical calculations.

The BCD for $tWSe_2$ bilayers has both $D_x$ and $D_y$ components because the only symmetry is $C_1$. From Ref. 4, if we apply a voltage $V_{/\!/}^{\omega}$ in the longitudinal direction, and measure the voltage $V_\perp^{2\omega}$ in the transverse direction, we can have:

$$\frac{V_\perp^{2\omega}}{\left(V_{/\!/}^{\omega}\right)^2} = \frac{e^3\tau}{2\hbar^2}\frac{\rho_{xx}\rho_{yy}(D_x\rho_{xx}cos\delta + D_y\rho_{yy}\sin\delta)}{\left(\rho_{xx}\cos^2\delta + \rho_{yy}\sin^2\delta\right)^2}$$

With $\delta$ is the angle between driving current and $x$ direction, $\tau = \frac{m^*}{ne^2\rho}$ is the scattering time obtained from the Drude formula, $\rho_{xx}$ and $\rho_{yy}$ is the resistance along zigzag and armchair direction respectively. In our disc-shaped sample, if we assume that the resistance is isotropic ($\rho_{xx} = \rho_{yy} = \rho_0$), we have

$$\frac{V_\perp^{2\omega}}{\left(V_{/\!/}^{\omega}\right)^2} = \frac{em^*}{2\hbar^2 p}\left(D_x cos\delta + D_y \sin\delta\right) \propto Dm^*.$$

Effectively, the total Berry curvature dipole $D = D_x cos\delta + D_y \sin\delta$.

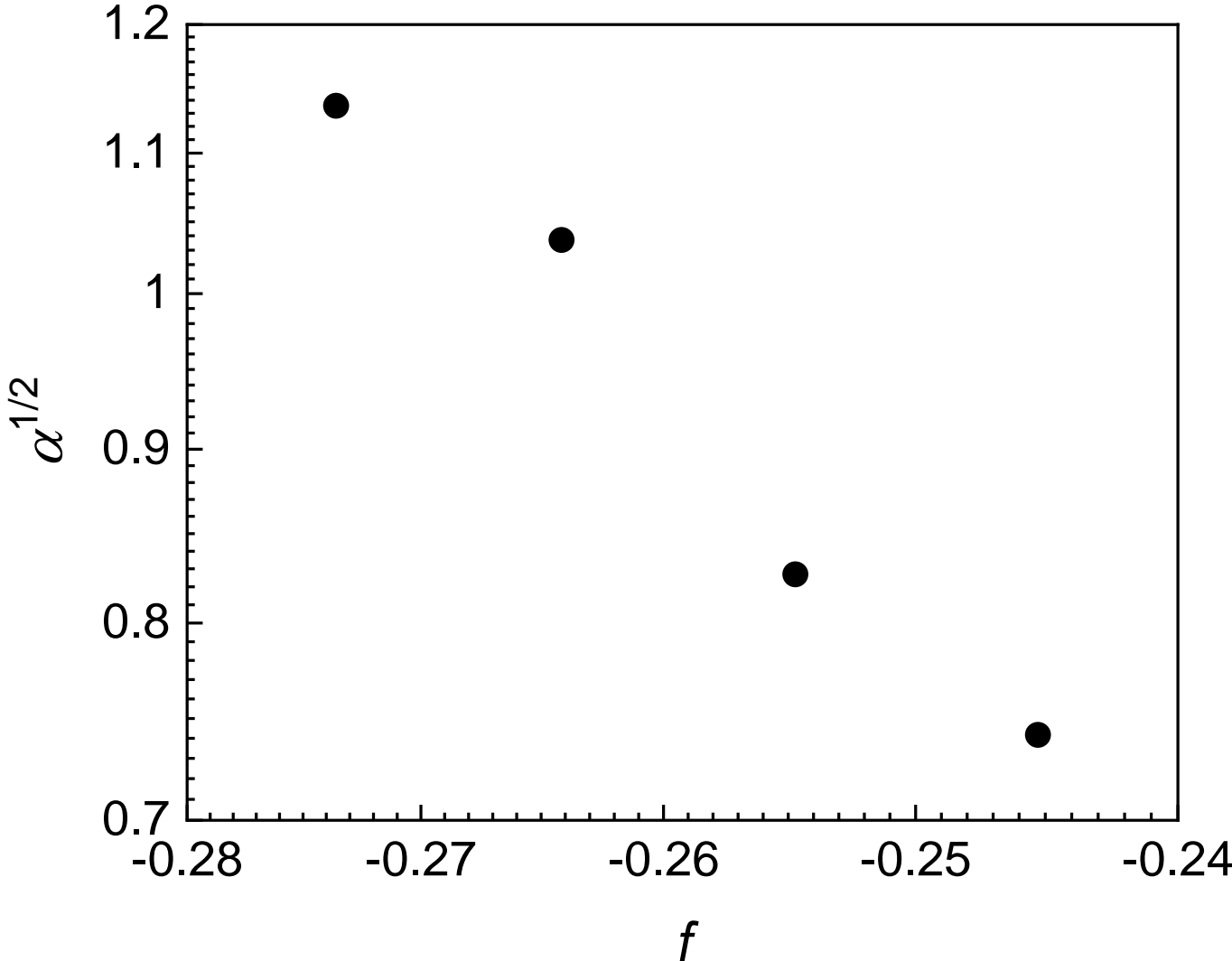

**Supplementary Fig. 5 | Filling dependence of $\alpha^{1/2}$ in a log–log plot.** The parameter $\alpha^{1/2}$, which proportional to the quasiparticle effective mass $m^*$ according to Kadowaki–Woods scaling[5], is enhanced as $f$ is approached from the metallic side.

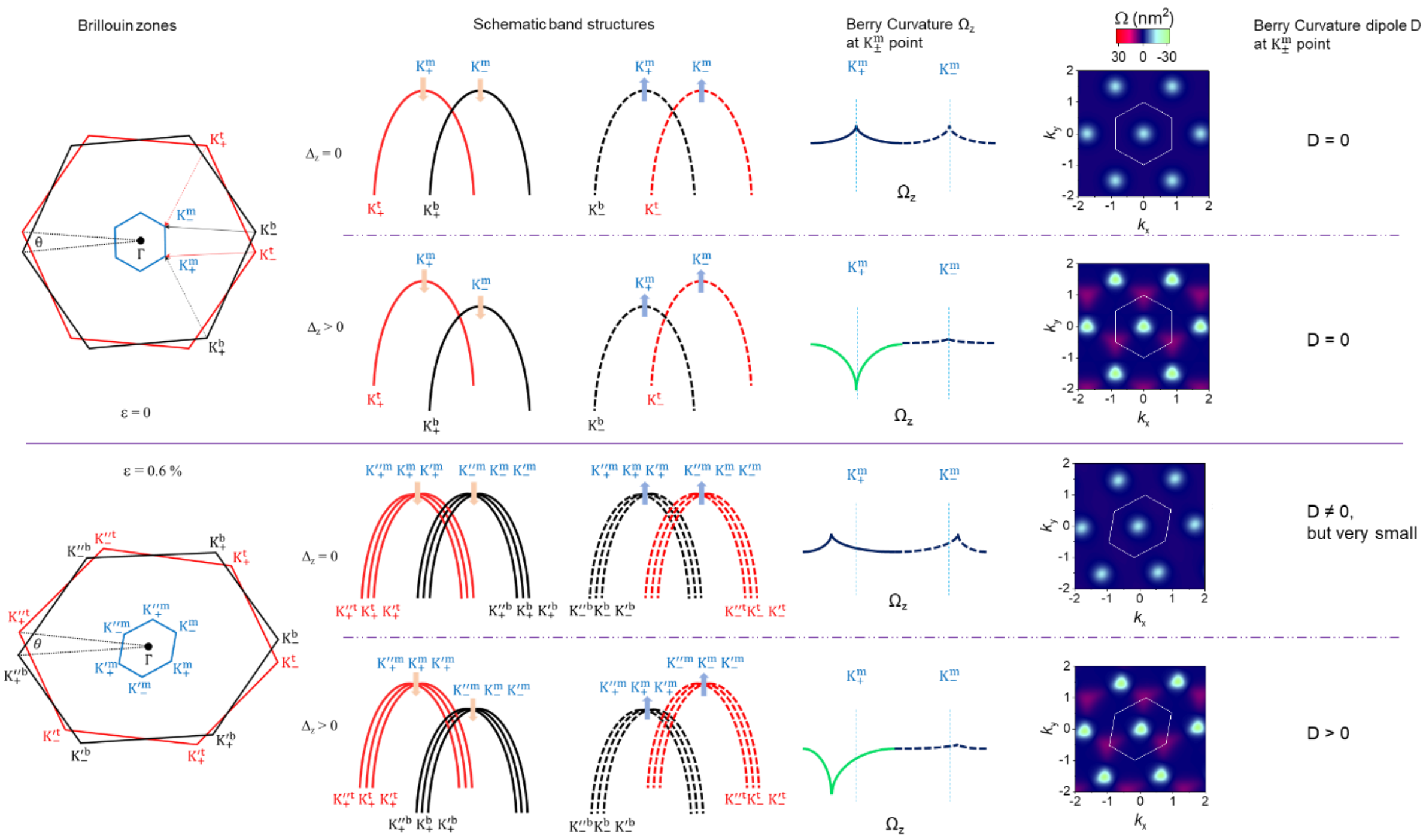

**Supplementary Fig. 6 | Strain and displacement electric field tunable Berry curvatures.**

1. $\epsilon = 0$, $\Delta_z = 0$ meV : $\Omega_z$ is small, $D = 0$ (three-fold symmetry forces the dipole to vanish)
2. $\epsilon = 0$, $\Delta_z$ =30 meV : $\Omega_z$ is large, $D = 0$ (three-fold symmetry forces the dipole to vanish)
3. $\epsilon = 0.6$ %, $\Delta_z = 0$ meV : $\Omega_z$ is small, $D \neq 0$ (the BC hotspot is small, so dipole is small)
4. $\epsilon = 0.6$ %, $\Delta_z = 30$ meV : $\Omega_z$ is large, $D > 0$ (the BC hotspot is large, so dipole is large)

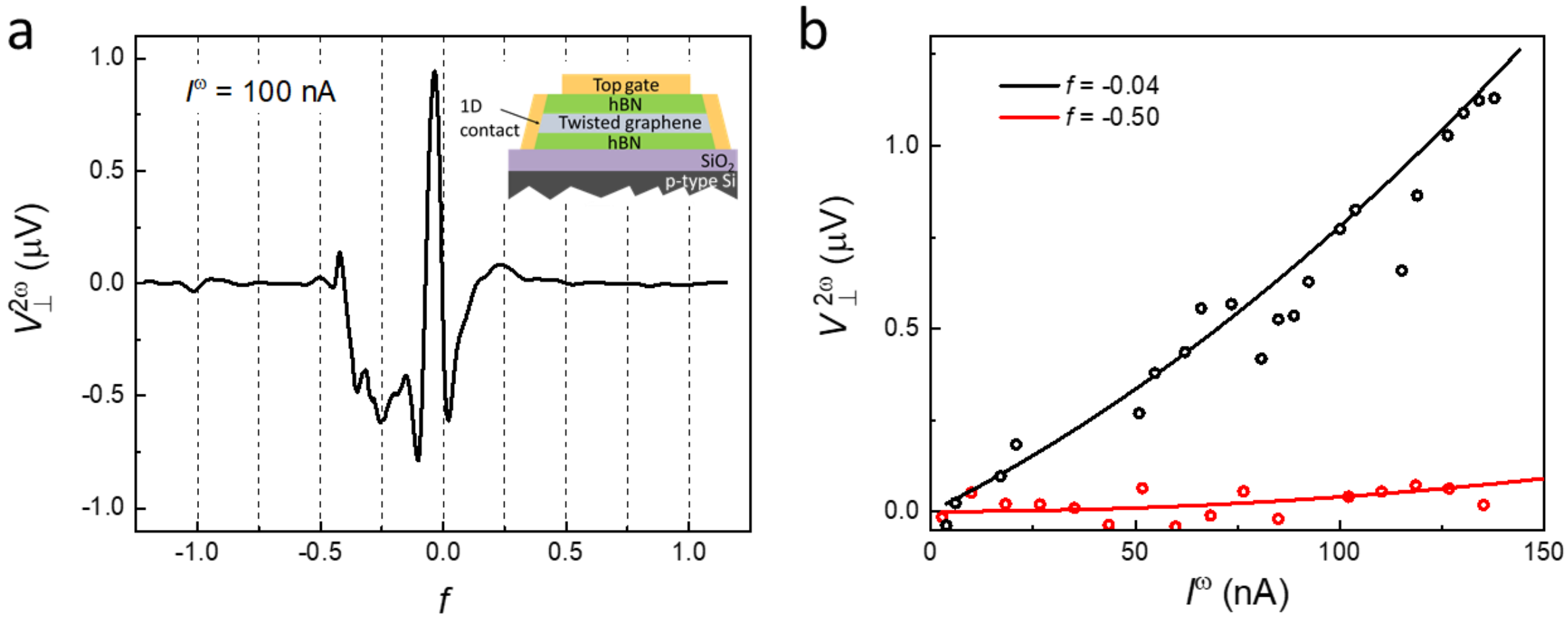

**Supplementary Fig. 7 | NHE in a 1.3° twisted bilayer graphene. a,** Filling dependent $V_{\perp}^{2\omega}$ measured at $T$=1.5 K. Inset: Schematic of device structure. The nonlinear Hall generation efficiency $\eta = 31\ \mathrm{V}^{-1}$ at $f = -0.04$ where $V_{\perp}^{2\omega}$ has its maximum value. **b,** $V_{\perp}^{2\omega}$ as a function of $I^{\omega}$ measured at different fillings. The dots are experimental data and the solid lines are parabolic fits of the data.

Though strain-induced nonlinear Hall signals can be detected in twisted bilayer graphene, the nonlinear Hall signal does not exhibit any giant behaviour near the half-filling. This is consistent with a theoretical prediction, saying that in twisted bilayer graphene the nonlinear Hall signal becomes apparent only for twist angles very close to the first magic angle ~1.1° (Ref. 6). Differently, there is no magic angle in twisted transition metal dichalcogenides: The bandwidth and BC hotspot can be tuned continuously along with the twist angle, demonstrating $tWSe_2$ to be a more accessible system for introducing and manipulating nonlinear Hall signals.

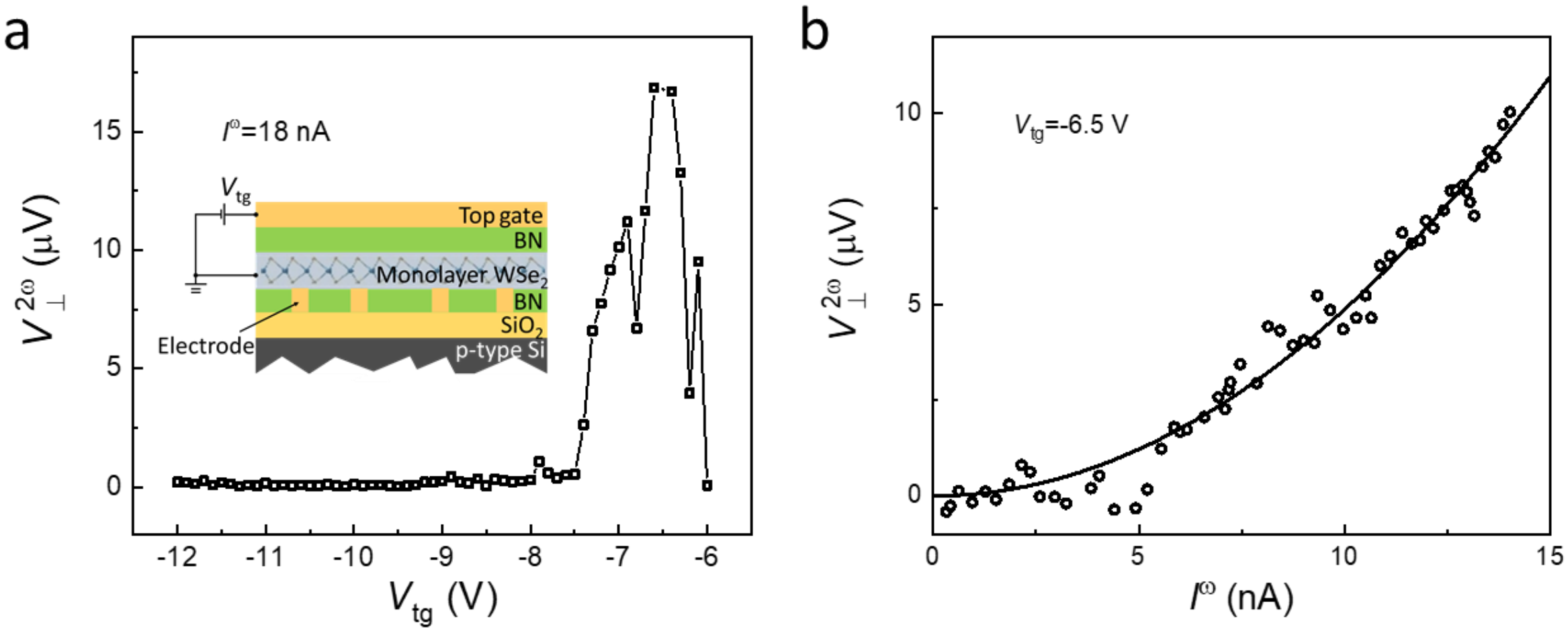

**Supplementary Fig. 8 | NHE in monolayer WSe₂. a,** Gate dependent $V_{\perp}^{2\omega}$ measured at $T$ =1.5 K, where $V_{\mathrm{tg}} = -6\mathrm{V}$ is the threshold voltage at which the Fermi level touches the band edge. Inset: Schematic of the device structure. **b,** $V_{\perp}^{2\omega}$ as a function of $I^{\omega}$ measured at $V_{\mathrm{tg}} = -6.5$ V. The dots are experimental data and the solid lines are parabolic fits of the data. The $V_{\perp}^{2\omega}$ signal in monolayer $WSe_2$, although weak, indicates the universal strain-induced three-fold symmetry breaking in the fabricated devices.

**Supplementary Table 1 | NHE in different 2D systems.**

| Materials | Mechanism | $\eta$ ($V^{-1}$) | $\frac{V_{\perp}^{2\omega}}{(I^{\omega})^2}$ ($VA^{-2}$) | $\frac{V_{\perp}^{2\omega}}{(I^{\omega})^2 R}$ ($VW^{-1}$) | Working Temperature (K) | Reference |
|---|---|---|---|---|---|---|
| $tWSe_2$ | Nonlinear Hall effect | 1190 | $7.1 \times 10^{18}$ | $9.2 \times 10^{10}$ | 1.5-40 | This work |
| Twisted bilayer graphene | | 37 | $8 \times 10^{7}$ | $5.4 \times 10^{4}$ | 1.5-80 | This work Supplementary Fig. 7 |
| Monolayer $WSe_2$ | | 0.017 | $5 \times 10^{10}$ | $2.9 \times 10^{4}$ | 1.5-90 | This work Supplementary Fig. 8 |
| Bilayer $WTe_2$ | | 2 | $2 \times 10^{8}$ | $2 \times 10^{4}$ | 10-40 | Ref [2] |
| Few layer $WTe_2$ | | 0.005 | 7.2 | 0.19 | 1.8-100 | Ref [7] |
| Corrugated graphene | | 6.25 | $1 \times 10^{8}$ | $2.5 \times 10^{4}$ | 3-15 | Ref [8] |
| $Bi_2Se_3$ | | 0.0000023 | 9 | 0.005 | 2-200 | Ref [9] |
| Strained monolayer $WSe_2$ | | 0.1 | $6.25 \times 10^{5}$ | 250 | 50-140 | Ref [10] |
| Graphene | FET-based multipliers | 0.0006 | - | - | 300 | Ref [11] |

## Supplementary Note 1 | The continuum model in $tWSe_2$.

The moiré bands are captured by the continuum Hamiltonian $H = \sum_\xi \int dr\, \psi_\xi^+(r)\, \widehat{H}_\xi(r)\psi_\xi(r)$ with

$$\widehat{H}_\xi(r) = \begin{pmatrix} \widehat{H}_{b,\xi} + U_b(r) & T_\xi(r) \\ T_\xi^+(r) & \widehat{H}_{t,\xi} + U_t(r) \end{pmatrix}$$

here the field operator $\psi_\xi^+ = (\psi_{b,\xi}^+, \psi_{t,\xi}^+)$. $\xi$ is the valley index. $\widehat{H}_{b(t),\xi}$ denotes the layer and valley dependent Hamiltonian and is given by $\widehat{H}_{l,\xi} = -\frac{\hbar^2}{2m^*}\left(\hat{k} - K_{\xi,l}^m\right)^2 - \frac{l\Delta_z}{2}$ where $l = b, t$ labels the bottom(top) layer, $m^*$ is the bare effective mass of valence band, $\Delta_z$ is the staggered layer potential generated by the vertical displacement field. $U_l(r),\ T(r)$ represent the intralayer and interlayer moiré potential respectively and are given by

$$U_l(r) = V \sum_{i=1,2,3} e^{i(g_i \cdot r + l\psi)} + h.c.$$

$$T_\xi(r) = w\left(1 + e^{-i\xi g_2 \cdot r} + e^{-i\xi(g_1+g_2)\cdot r}\right)$$

where the moiré reciprocal lattice vectors $g_i = \frac{4\pi}{\sqrt{3}L_M}\left(cos\frac{2(i-1)\pi}{3}, sin\frac{2(i-1)\pi}{3}\right)$. The model parameters are $(m^*, w, V, \psi) = (0.44m_e, 18meV, 5meV, 91^\circ)$ [12, 13]. The staggered layer potential in $tWSe_2$ can be calculated roughly by $\Delta_z = \frac{C_{bg}V_{\mathrm{bg}} - C_{tg}V_{\mathrm{tg}}}{C_{\mathrm{Q}}}$, where $C_{\mathrm{Q}} = \frac{2e^2m^*}{\pi\hbar^2} \approx 10^{-4}\ \mathrm{F\ cm^{-2}}$ is the quantum capacitance of monolayer $WSe_2$.

The strained twisted $WSe_2$ is modelled by introducing a strain along the $\varphi$ direction with a strength $\epsilon$ on the bottom layer. The strain tensor $\boldsymbol{\varepsilon}$ can be written as $\boldsymbol{\varepsilon} = \epsilon\begin{pmatrix} \cos^2\varphi - \upsilon\sin^2\varphi & (1+\delta)\cos\varphi\sin\varphi \\ (1+\delta)\cos\varphi\sin\varphi & \sin^2\varphi - \upsilon\cos^2\varphi \end{pmatrix}$, where $\delta = 0.19$ is the Poisson ratio[14] of $WSe_2$. Under this strain, the Dirac point for monolayer TMDC is shifted to $D_\xi = (I - \varepsilon)K_\xi - \xi A$ with the effective gauge field $A = \frac{\sqrt{3}}{2a_0}\beta\left(\epsilon_{xx} - \epsilon_{yy}, -2\epsilon_{xy}\right)$. $\beta$ is adopted as 2.30 in our calculation according to previous first principle calculation for strained $WSe_2$ (Ref. 15). Then the continuum Hamiltonian of this strained $WSe_2$ can be obtained as $\widehat{H}_{b,\xi} = -\frac{\hbar^2}{2m^*}\left(\hat{k} - D_\xi^m\right)^2 - \frac{\Delta_z}{2}$.

The Berry curvature dipole can be evaluated by:

$$D = -\int dk_x dk_y \sum_{\alpha,\xi} v_{\alpha k,\xi}^{x(y)}\, \Omega_{\alpha k,\xi}\delta_F(E_\alpha - E_F)$$

where $\alpha, \xi$ are band and valley index. $\Omega_{\alpha k,\xi} = i\langle\partial_{\boldsymbol{k}}u_{\alpha k,\xi}| \times |\partial_{\boldsymbol{k}}u_{\alpha k,\xi}\rangle$ is the BC, and $|u_{\alpha k,\xi}\rangle$ is the Bloch wavefunction obtained from the continuum model. $v_{\alpha k,\xi}^{x(y)}$ is the band velocity and $\delta_F(E_\alpha - E_F)$ is a delta-function centered at $E_F$. Detailed calculations of the BCD for strained $tWSe_2$ can be found in reference 16.